\documentclass[epsfig,psfig,aps,twocolumn,prb,showpacs]{revtex4-1}
\usepackage{amsfonts,amsmath,bbm,wasysym}
\usepackage{amsfonts,dsfont}
\usepackage[english]{babel}
\usepackage{float}
\usepackage{lipsum}
\usepackage{graphicx}
\usepackage{epsfig}
\usepackage{relsize}
\usepackage[colorlinks=true,citecolor=red,urlcolor=blue]{hyperref}
\usepackage{soul}

 \usepackage{bm}

\begin{document}

\title{Fragility of the antichiral edge states under disorder\\
}
\author{Marwa Manna\"i$^{1,2}$}
\author{Eduardo V. Castro$^{3,4}$}
\author{Sonia Haddad$^{2,5,6}$}
\affiliation{
$^1$Center for Quantum and Topological Systems, NYUAD Research Institute, New York University Abu Dhabi, UAE\\
$^2$Laboratoire de Physique de la Mati\`ere Condens\'ee, Facult\'e des Sciences de Tunis, Universit\'e Tunis El Manar, Campus Universitaire 1060 Tunis, Tunisia\\
$^3$ Centro de F\'isica das Universidades do Minho e Porto, Departamento de F\'isica e Astronomia, Faculdade de Ci\^encias, Universidade do Porto, 4169-007 Porto, Portugal\\
$^4$ Beijing Computational Science Research Center, Beijing 100084, China\\
$^5$ Max Planck Institute for the Physics of Complex Systems, N\"othnitzer Strasse 38, Dresden 01187, Germany\\
$^6$ Institute  for  Theoretical  Solid  State  Physics,  IFW  Dresden, Helmholtzstr.   20,  01069  Dresden,  Germany
}
\date{\today}

\begin{abstract}
Chiral edge states are the fingerprint of the bulk-edge correspondence in a Chern insulator.
Co-propagating edge modes, known as antichiral edge states, have been predicted to occur in the so-called modified Haldane model describing a two-dimensional semi-metal with broken time reversal symmetry. 
These counterintuitive edge modes are argued to be immune to backscattering and extremely robust against disorder. Here, we investigate the robustness of the  antichiral edge states in the presence of Anderson disorder. By analysing different localization parameters, we show that, contrary to the general belief, these edge modes are fragile against disorder, and can be easily localized. Our work provides insights to improve the transport efficiency in the burgeoning fields of antichiral topological photonics and acoustics.
%
\end{abstract}

\maketitle

\section {Introduction} The hallmark property of topological materials is the the occurrence of topologically protected edge channels propagating at the boundaries of the bulk structure ~\cite{Hasan-Rev,Qi-Rev,Bansil}.
Regarding their robustness against disorder, these states can support dissipationless current flow. The flagship edge states are the chiral modes of the Haldane Chern insulator ~\cite{Haldane} describing a spinless electronic system on a honeycomb lattice, where time reversal symmetry (TRS) is broken by complex hopping integrals between next nearest-neighboring (NNN) atoms. The latter are characterized by a complex phase $\Phi$, which results into staggered magnetic fluxes having the same configuration in the two honeycomb sublattices.
Different systems~\cite{AQHE-rev,Haldane2,Haldane3,photonic,Niu,Wang,Yang,Khan,Ni,kagome,Kee,Chern-gr,Chern-weyl} have been proposed to host chiral edge states  which have been observed in photonic crystals ~\cite{Wang09}, optical lattices ~\cite{cold}, acoustic systems ~\cite{acoustic}, thin film of magnetic topological insulators ~\cite{Rosen}, magnetic Weyl semimetals ~\cite{Chern-weyl} and in  nanomechanical graphene ~\cite{gr}.

By flipping the sign of the complex phase of the NNN integrals in one sublattice, Colom\'es and Franz~\cite{Franz} obtained the so-called modified Haldane model (mHM) describing a semi-metal with broken TRS resulting from a valley-dependent pseudo-scalar potential that offsets the Dirac point energies. In a zizgag ribbon geometry, the mHM gives rise to co-propagating edge modes, known as the antichiral (AC) edge states, connecting the oppositely shifted Dirac points. Regarding the ungapped spectrum of the model, the AC modes are counterbalanced by an equal number of gapless bulk states which propagate in the opposite direction.

AC edge states are expected to be implemented in transition-metal dichalcogenides ~\cite{Franz}, exciton-polariton systems~\cite{Mandal,Mandal22}, gyromagnetic photonic crystals~\cite{Chen20}, acoustic resonators~\cite{JAP21}, twisted van der Waals multilayers ~\cite{Denner}, combined systems of Haldane model~\cite{Cheng}, Heisenberg ferromagnets on honeycomb lattice~\cite{Bhowmick20}, and Floquet lattices~\cite{Floquet}.

Recently, there has been a growing interest in the topological properties of the antichiral edge states and their possible applications~\cite{Mannai,Roche,supra,Chen,AC-mag,helical,coexist}. It has been shown that a Bernal stacked bilayer of the mHM gives rise to a Chern insulator with a Chen number $C=0, \,\pm 1,\, \mathrm{or}\, \pm 2$ ~\cite{Marwa23}. One-way bulk states were predicted in a strip of alternately stacked modified Haldane lattices with opposite complex phases~\cite{heteroHM}.

The experimental realization of the AC edge states has been reported in a microwave-scale gyromagnetic photonic crystal~\cite{Zhou,photonic22} and in electrical circuits ~\cite{Yang21}. AC surface states, a 2D extension of the one-dimensional AC edge states of the mHM, have also been observed in photonic crystal~\cite{Weyl,surface-AC}.\newline
The outcomes of these studies point towards the promising applications of AC edge states in topological photonics.  But, are AC edge states topologically robust? \newline
As the mMH is gapless, the AC edge states cannot be protected by a bulk topological invariant. However, they were predicted to be robust against disorder in the same way as the zero energy edge modes of a graphene zigzag ribbon~\cite{Franz}.\

In this work, we address the robustness of the AC edge states against Anderson disorder. We first compute, based on the coupling matrix approach, the winding number of a one-dimensional (1D)-reduced mHM and analyze its behavior under an on-site disorder. 
Then, we focus on strips described by the Haldane model (HM) and the mHM. We compute, using the transfer matrix method, the localization lengths of, respectively, the chiral and the AC edge states. Our results show that, contrary to the chiral states, the AC edge modes are not robust and can be easily localized by defects which mix them with their counterbalancing bulk modes.\\
The paper is organized as follow. In section II, we derive the 1D description of the mHM which is mapped to an extended Su-Schrieffer-Heeger (SSH) model with momentum dependent hopping integrals. We then study the behavior of the corresponding winding number and the inverse participation ratio (IPR) in the presence of Anderson type disorder. In section III, we compute the localization lengths of zigzag ribbons of the HM and the mHM with different widths. We benchmark the behaviors of the associated localization lengths under different disorder amplitudes and concentrations. The concluding section IV summarizes our results.
\section{1D-reduced modified Haldane model}
\subsection{Modified Haldane model: a brief review} 
%
 \begin{figure}[hpbt] 
\centering
\includegraphics[width=1\columnwidth]{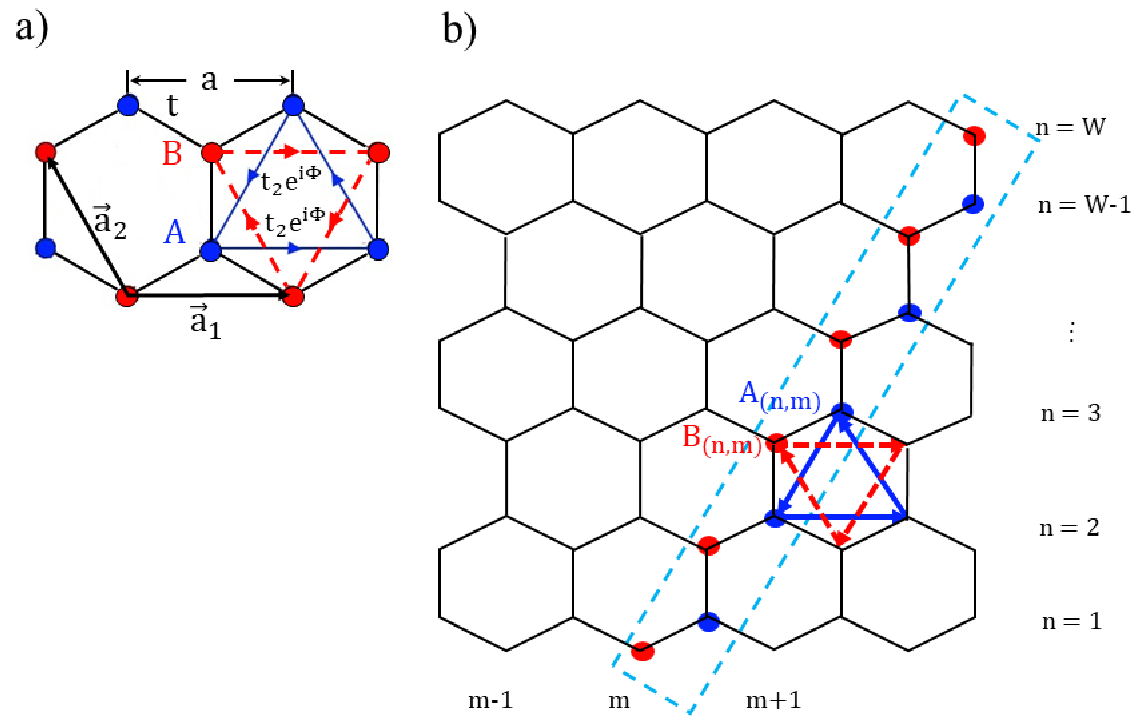}
\caption{(a) Pattern of the NNN hopping processes of the mHM. The blue and red arrows indicate the directions along which the electron gains a phase $\Phi$. $t$ and $t_2$ denote, respectively, the NN and NNN hopping integrals. $\left(\mathbf{a}_1,\,\mathbf{a}_2\right)$ is the lattice basis and $a$ is the lattice parameter. (b) The mHM on a zigzag nanoribbon structure of width $W$. A lattice site is expressed as $\mathbf{R}_{m,n}=m\mathbf{a}_1+n\mathbf{a}_2$. The dashed  rectangle delimits the ribbon unit cell used to define the reduced 1D-mHM.}
\label{structure}
\end{figure}
We consider the mHM  where the NNN hopping integrals have a complex phase with opposite signs in the two honeycomb sublattices (Fig.~\ref{structure} (a)). The corresponding Hamiltonian is
\begin{eqnarray}
H=t \sum_{\langle i,j\rangle }c_i^{\dagger} c_j+ t_2\sum_{\langle\langle i,j\rangle\rangle } e^{i \Phi_{i,j}} c_i^{\dagger} c_j,
\label{mHM1}
\end{eqnarray}
where $c_i$ is the annihilation operator of a spinless electron on an atom of the honeycomb lattice, $t$ and $t_2$ are the hopping integrals between, respectively, the nearest neighboring (NN) and the next nearest neighboring (NNN) atoms, 
and $ \Phi_{i,j}=\Phi$ ($-\Phi$) for NNN hopping processes along (in the opposite direction to) the pattern shown in Fig.~\ref{structure} (a). Without loss of generality, we hereafter take $t_2=0.1t$ and $\Phi=\frac{\pi}2$~\cite{Franz}.\

\begin{figure}[hpbt] 
\centering
\includegraphics[width=0.8\columnwidth]{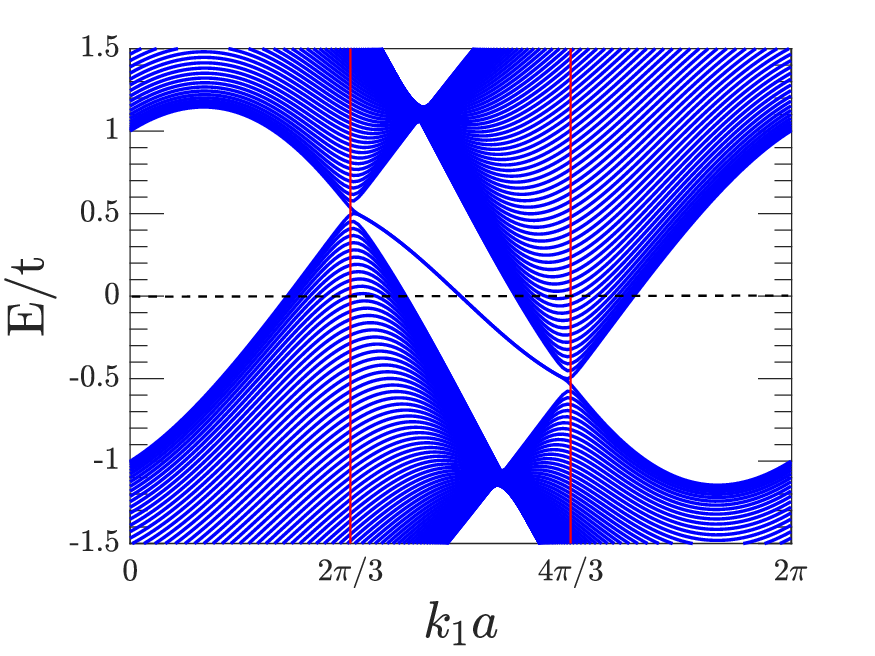}
\caption{Energy spectrum of a mHM on a ribbon of a width $W=100$. The red lines indicate the positions of the Dirac points connected by the co-propagating AC edge states. Calculations are done for $t_2=0.1t$, and $\Phi=\frac{\pi}2$.}
\label{energy}
\end{figure}

The corresponding band structure, in a zigzag nanoribbon, shows a semi-metallic behavior with co-propagating gapless AC edge states which connect the oppositely shifted Dirac points (Fig.~\ref{energy}).

Based on the conductance behavior of the mHM in a nanoribbon, Colom\'es and Franz~\cite{Franz} argued that the AC edge modes are robust against on-site disorder. 
By increasing the ribbon length, the conductance reaches a value of 2 (in units of $e^2/h$), which was ascribed to the two AC edge modes crossed by the Fermi level. This behavior is in contrast with that observed in graphene zigzag ribbon~\cite{Gr-cond}, where the conductance is found to collapse. According to the authors~\cite{Franz}, the AC edge states remain delocalized even at large impurity concentrations, which was attributed to their large localization length compared to that of the bulk modes~\cite{Franz}. However, the authors pointed out that the behavior of the counter-propagating modes, accompanying the AC edge states, is unclear. These states are found to be concentrated along the sample boundaries instead of being fully extended over the width. To get insights on the topological character of the AC edge states, the authors~\cite{Franz} analyzed the topology of the zero energy modes of the zigzag graphene ribbon which are at the origin of the AC modes of the mHM. They defined a 1D-reduced graphene Bloch Hamiltonian by fixing the momentum component along the zigzag direction of the ribbon, and derived the corresponding winding number~\cite{Franz}. The latter shows a quantized nonzero value, indicating that the pristine zigzag edge modes of graphene are topologically protected in the same way as the chiral symmetry class AIII of topological insulators~\cite{AZ}. Since the Bloch Hamiltonian of the mHM, (see next section), has the same eigenstates as graphene, Colom\'es and Franz~\cite{Franz} concluded that the AC edge states have the same topological properties as the zigzag edge modes of graphene. As a consequence they are expected to be robust against disorder.\newline
Such conclusion does not take into account the presence of the counter-propagating bulk modes which can couple to the AC edge states in the presence of defects ~\cite{Mandal22}. \

To address the topology of the AC edge states, we derive, in the following, the effective 1D-mHM and the corresponding winding number based on the approach of Ref.~[\onlinecite{SSH-HM}], where the authors showed that the 2D HM can be mapped into an extended SSH model~\cite{SSH}. We then analyse the behavior of the winding number of the mHM in the presence of Anderson type disorder~\cite{Phil}.
\subsection{1D-modified Haldane model: winding number}
To obtain the reduced 1D-Hamiltonian of the mHM, we start by performing a partial Fourier transformation of the real space Hamiltonian (Eq.~\ref{mHM1}), with the respect to the $x$ component, along the zigzag ends, of the atomic positions. The annihilation operator $c_i\equiv c_{\alpha}\left(\mathbf{R}_{m,n}\right)$ of an atom $\alpha$ ($\alpha=A,\,B$) belonging to the unit cell $\mathbf{R}_{m,n}=m \mathbf{a_1}+n \mathbf{a_2}$, can be written as
\begin{eqnarray}
c_i=\frac 1{\sqrt{N_1}}\sum_{k_1} e^{i k_1 x_{m,n}^{\alpha}} c_{\alpha,n}(k_1),
\label{TFc}
\end{eqnarray}
where ($\mathbf{a_1},\mathbf{a_2}$) is the graphene basis (Fig.~\ref{structure}), $x_{m,n}^{\alpha}$ is the position of the atom $\alpha$ in the unit cell $\mathbf{R}_{m,n}$, $k_1=\mathbf{k}\cdot \frac{\mathbf{a_1}}a$ is the momentum component along $\mathbf{a_1}$ axis, and $a$ is the graphene lattice parameter.
This Fourier transform turns out to write the Hamiltonian in the so-called basis II~\cite{Bena,JN}, or within the convention II~\cite{Vanderbilt}, which takes into account the atom positions in the unit cells. This basis should be used to derive the physical properties of the system~\cite{JN}, which are the hopping integrals of the $k_1$-dependent Hamiltonian $H(k_1)$ in terms of which is expressed the Hamiltonian of Eq.~\ref{mHM1} as
\begin{eqnarray}
H&=&\sum_{k_1} H(k_1),\\
H(k_1)&=&J_1 \sum_{n} c_{A,n+1}^{\dagger} c_{B,n}+
J_1^{\prime} \sum_{n} c_{A,n}^{\dagger} c_{B,n}\nonumber\\
&+&J_2 \sum_{n,\alpha} c_{\alpha,n+1}^{\dagger} c_{\alpha,n}
+\mu(k_1) \sum_{n,\alpha} c_{\alpha,n}^{\dagger} c_{\alpha,n}+h.c.,\nonumber\\
\label{mHM2}
\end{eqnarray}
where $J_1=t$, $J_1^{\prime}=2t \cos\left(\frac a2k_1\right)$, $J_2=2t_2\cos\left(\frac a2k_1-\Phi\right)$ and $\mu(k_1)=2t_2\cos\left(a k_1+\Phi\right)$.

$H(k_1)$ is reminiscent of the Hamiltonian of the extended SSH model~\cite{SSH-HM} with an equal onsite chemical potential $\mu_A=\mu_B=\mu(k_1)$.\

To derive the winding number of the reduced 1D Hamiltonian (Eq.~\ref{mHM2}), we perform a second Fourier transform with respect to the atomic position along the $\mathbf{a_2}$ direction as
\begin{eqnarray}
c_{\alpha,n}(k_1)=\frac 1{\sqrt{N_2}}\sum_{k_2} e^{i k_2 X_{n}} c_{\alpha,k_1}(k_2),
\end{eqnarray}
where $X_n=na$ is the position of the unit cell, $N_2$ is the number of unit cells along $\mathbf{a_2}$, and $k_2=\mathbf{k}\cdot\frac{\mathbf{a_2}}a $. 
Here, we adopted the so-called basis I description~\cite{Bena,JN,Vanderbilt} which gives rise, unlike the basis II, to a periodic Bloch Hamiltonian which will be used to define the winding number~\cite{Cayssol}. \newline
Carrying out the Fourier transform, we obtain the Hamiltonian $H_{k_1}(k_2)$ of the reduced 1D mHM as
\begin{eqnarray}
H_{k_1}(k_2)=d_{0,k_1}(k_2)\sigma_0+\mathbf{d}_{k_1}(k_2)\cdot\mathbf{\sigma},
\label{mHM3}
\end{eqnarray}
where $\mathbf{\sigma}$ are the sublattice Pauli matrices, $\sigma_0$ is the $2\times 2$ identity matrix, $d_{0,k_1}(k_2)=J_2\cos\left(k_2a\right)+\mu(k_1)$, $d_{x,k_1}(k_2)=J_1^{\prime}+J_1\cos\left(k_2a\right)$, $d_{y,k_1}(k_2)=J_1\sin\left(k_2a\right)$, and $d_{z,k_1}(k_2)=0$.\newline
The corresponding dispersion relation is $E_{k_1}(k_2)=d_{0,k_1}(k_2)\pm |\mathbf{d}_{k_1}(k_2)|$, which describes two bands separated by a gap that closes at the Dirac points $K_{\xi}=\left(k_1=\xi\frac{2\pi}{3a},k_2=\frac{\pi}{a}\right)$ if $|J_1^{\prime}|=J_1$. Here $\xi$ is the valley index on which depends the scalar term $d_{0,k_1}(k_2)$ responsible of the offset of the Dirac point energies.\

The reduced 1D-mHM given by Eq.~\ref{mHM3} breaks TRS due to the complex phase of the NNN hopping integrals (Eq.~\ref{mHM2}). Chiral and particle-hole symmetries are also broken since $\sigma_z H_{k_1}(k_2)\sigma_z\neq -H_{k_1}(k_2)$ and $\sigma_z H_{k_1}^{\ast}(k_2)\sigma_z\neq- H_{k_1}(-k_2)$, respectively. \

In the absence of NNN hopping processes, $H_{k_1}(k_2)$ reduces to the standard SSH model~\cite{SSH}, with effective NN hopping integrals $J_1$ and $J^{\prime}_1$ (Eq.~\ref{mHM2}), characterized by its winding number
\begin{eqnarray}
 \nu_{k_1}&=&\frac i{\pi}\int dk_2\langle u_{k_1}(k_2)|\partial_{k_2} u_{k_1}(k_2)\rangle\nonumber\\
 &=&\frac 1{2\pi} \int_{0}^{2\pi}dk_2 \frac{d\Phi_{k_1}(k_2)}{dk_2},
 \label{nu}
\end{eqnarray}
where $ |u_{k_1}(k_2)\rangle=\frac 1{\sqrt{2}}\left(e^{i\Phi_{k_1}(k_2)}, \pm 1\right)^{\dagger}$ and $\Phi_{k_1}(k_2)=\mathrm{Im} \left[\ln\left(J_1^{\prime}+J_1 e^{ik_2a}\right)\right]$.\

The winding number, given by Eq.~\ref{nu}, can be also ascribed to the reduced 1D-mHM Hamiltonian (Eq.~\ref{mHM3}) which differs from the SSH Hamiltonian by a diagonal term. The latter does not affect the eigenstates on which depends the winding number. It comes out that the 1D-mHM has a winding number satisfying~\cite{SSH}
\begin{eqnarray}
 |\nu_{k_1}|=\left\{ 
 \begin{array}{cc}
  1,& \; \mathrm {if}\; |J_1^{\prime}|<J_1,\\
  0, &\; \mathrm {if} \; |J_1^{\prime}|>J_1,\\
  \mathrm {undefined}& \;\mathrm {if} |J_1^{\prime}|=J_1.
 \end{array}
 \right.
 \label{nu-cond}
\end{eqnarray}
The pristine 1D-mHM is then topologically non trivial if $\frac{2\pi}{3a}<k_1<\frac{4\pi}{3a}$, where the limiting values correspond to the Dirac point positions along the zigzag ribbon direction. This result is in agreement with that found in Ref.~[\onlinecite{Franz}] where the authors ascribed the topology of the AC edge states to the non-vanishing winding number of the zigzag graphene ribbon in the absence of NNN term. \

To address the topological protection of the AC edge states, one needs to study the behavior of the winding number $\nu_{k_1}$ of the 1D-mHM under disorder.
\subsection{Disordered 1D-modified Haldane model: winding number} 
\begin{figure}[hpbt] 
\centering
\includegraphics[width=1\columnwidth]{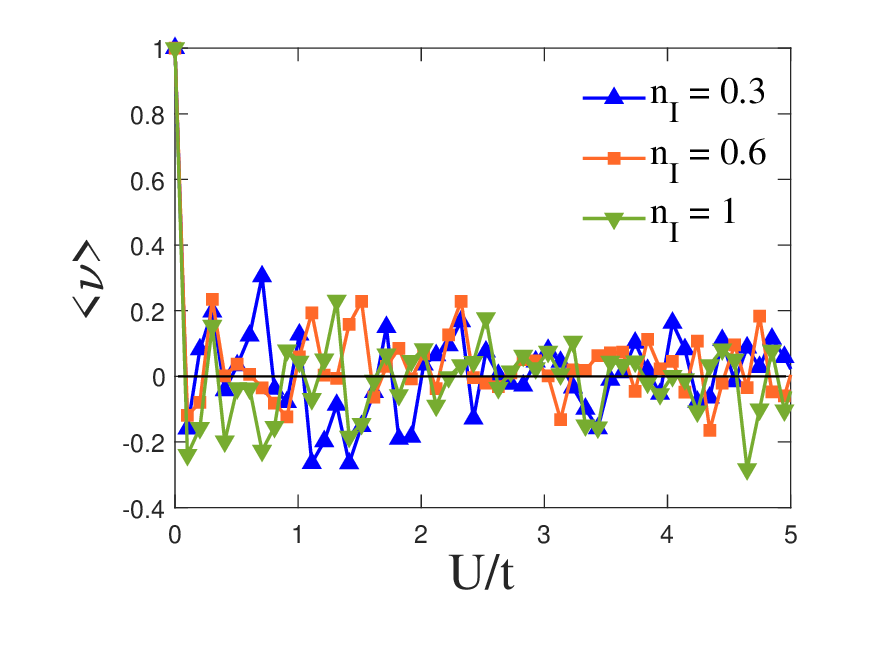}
\caption{Averaged winding number of the reduced 1D-mHM as a function of the normalized disorder amplitude $U/t$ for different disorder concentrations $n_I$. Calculations are done for $t_2=0.1t$, $\Phi=\frac{\pi}2$, $k_1=\frac{7\pi}{6a}$ and a ribbon width $W = 75$.
The average is taken over $100$ disorder configurations.}
\label{fig-nu}
\end{figure}

We consider the 1D-mHM in the presence of an onsite Anderson disorder which modifies the last term in the Hamiltonian $H(k_1)$ (Eq.~\ref{mHM2}) as
\begin{eqnarray}
\mu(k_1)\sum_{n,\alpha} c_{\alpha,n}^{\dagger} c_{\alpha,n}\longrightarrow
\sum_{n,\alpha}\mu_n(k_1) c_{\alpha,n}^{\dagger} c_{\alpha,n},
\label{mHM4}
\end{eqnarray}
where $\mu_n(k_1)=\mu(k_1)+Uw_n$, $U$ is the disorder amplitude, and $w_n$ is a uniform random number $w_n\in\left[-\frac12,\frac 12\right]$.
In the presence of disorder, the winding number cannot be derived analytically since the integration over the Brillouin zone is no more possible regarding the broken translational symmetry. The winding number can be computed numerically using the complex matrix method~\cite{CM}, where the momentum component $k_2$, in Eq.~\ref{nu}, is replaced by a phase twist $\theta$ of the real space single-particle wavefunction.\newline
For a long chain ($L\gg W$) (Fig.~\ref{structure}), one can use the numerical approach of Ref.~[\onlinecite{CM}] which significantly reduces the computational time by carrying out  the calculations in the momentum space with twisted boundary conditions.
Based on this approach, we computed the averaged winding number $\langle\nu_{k_1}\rangle$ of the disordered 1D-mHM described by the Hamiltonian $H(k_1)$ (Eq.~\ref{mHM2}) with the disorder potential given by Eq.~\ref{mHM4}. The results are
depicted in Fig.~\ref{fig-nu}.\newline

In the pristine system ($U=0$), $\langle\nu_{k_1}\rangle$ is quantized to 1, but it drastically vanishes under disorder. The strong suppression of the quantization occurs at relatively weak disorder amplitude ($U< 0.2t$) and is particularly independent of the impurity concentration. This behavior shows that the winding number cannot be taken as a probe for the topological properties of the disordered 1D-mHM since the chiral symmetry, defining the 1D topological insulator class, is broken under Anderson disorder.\newline
It is worth to stress that the Chern number quantization of the HM is found to be a robust feature~\cite{CM}. The corresponding plateau at $C=1$ persists up to a strong disorder amplitude ($U\sim 4t$), which expresses the extreme robustness of the chiral edge states compared to the AC edge channels.

\subsection{Disordered 1D-modified Haldane model: IPR} 

Besides the winding number, the disorder-induced localization of the AC edge states can be characterized by the inverse participation ratio (IPR) defined for an eigenstate $|\varphi_n\rangle$ of the Hamiltonian by $\text{IPR}\left(|\varphi_n\rangle\right)=\sum_i|\varphi\left(\mathbf{r}_i\right)|^4$, where $\varphi\left(\mathbf{r}_i\right)$ is the amplitude of the eigenstate in the site $|\mathbf{r}_i\rangle$.\newline
In the thermodynamic limit, the IPR of a delocalized state vanishes, while it remains finite for a localized state and reaches 1 for a state completely localized on one lattice site.\

In the SSH model, the IPR of the bulk states increases with increasing disorder amplitude, which indicates a disorder induced localization~\cite{SSH-IPR}. However, the SSH edge states are found to have a decreasing IPR reflecting a defect-induced broken chiral symmetry, which delocalize the edge modes~\cite{SSH-IPR}. \

\begin{figure}[hpbt] 
\includegraphics[width=0.7\columnwidth]{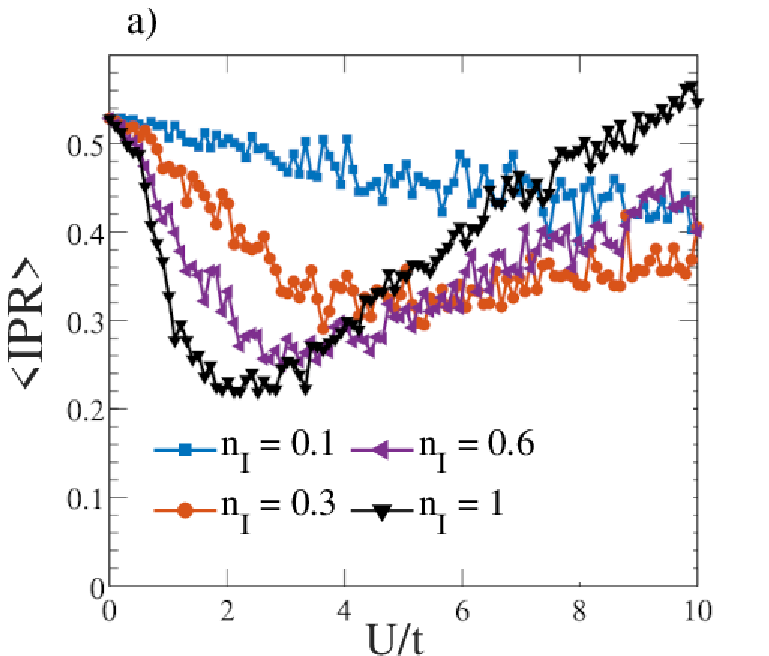}
\includegraphics[width=0.7\columnwidth]{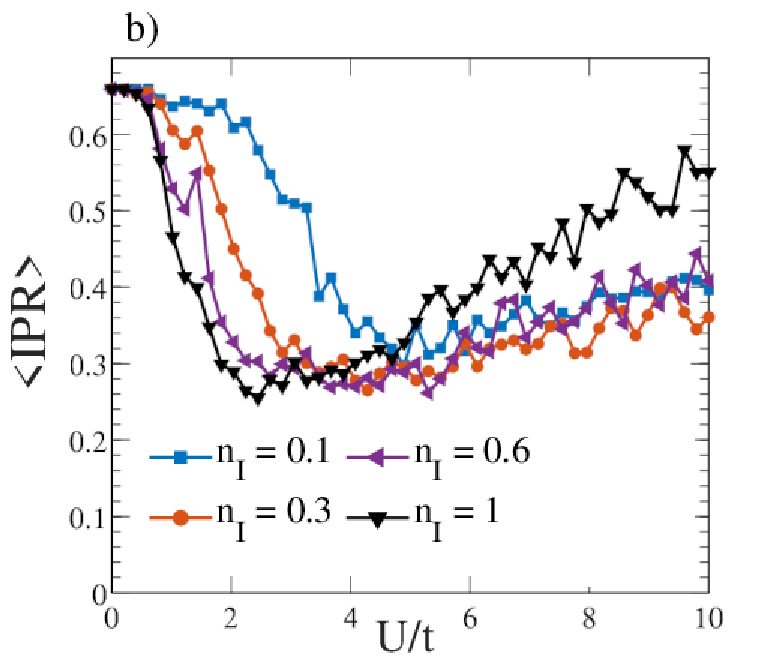}
\caption{Averaged IPR of the mid-gap edge states as a function of the normalized disorder amplitude $U/t$ at different impurity concentrations $n_I$ for the 1D reduced (a) mHM and (b) HM. The IPR is averaged over $100$ random configurations. Calculations are done for $t_2 = 0.1t$, $\Phi=\pi/2$, $k_1=\frac{7\pi}{6a}$ and a chain of $n = 150$ sites (Fig.~\ref{structure}(b)). }
\label{IPR}
\end{figure}

Figure~\ref{IPR} represents the IPR as a function of the disorder amplitude of the mid-gap states of the reduced 1D-mHM and 1D-reduced HM at different impurity concentrations. The IPR are averaged over 100 disorder configurations.
At low concentrations ($n_I\sim0.1$), Fig.~\ref{IPR} (a) shows that the IPR of the AC edge states decreases slowly with increasing disorder amplitude, which corresponds to the disorder-induced delocalization resulting, as in the SSH model, from the chiral symmetry breaking.
At higher concentrations, the AC edge states are delocalized by defects up to a critical value of the disorder amplitude above which they undergo an Anderson localization. The localization regime is rapidly reached by increasing $n_I$ since the number of defected sites increases~\cite{Franz}. It comes out that the decrease of the IPR of the AC edge modes does not reflect the robustness of these states.\

In the case of the HM (Fig.~\ref{IPR} (b)), the chiral symmetry is already broken in the pristine system and the drop of the IPR indicates the delocalization of the chiral edge states which expand on the sites closer to the boundaries before mixing with the bulk edge states above a critical disorder amplitude. The latter is shifted towards smaller values as the impurity concentration increases. The chiral edge states remain, then, localized around the system ends at relatively weak disorder amplitudes, which is an indicator of a protected bulk topology.\

Actually, the IPR behavior should be taken with a grain of salt since non-topological states can have IPR exhibiting similar behavior as topological modes ~\cite{IPR}. To avoid confusing conclusions, we compute, in the following, the localization lengths of the edge states of the 2D mHM and the 2D HM.\\
\section{ 2D-modified Haldane model: Localization lengths} 
\begin{figure*}[hpbt] 
\centering
$\begin{array}{ccc}
\includegraphics[width=0.7\columnwidth]{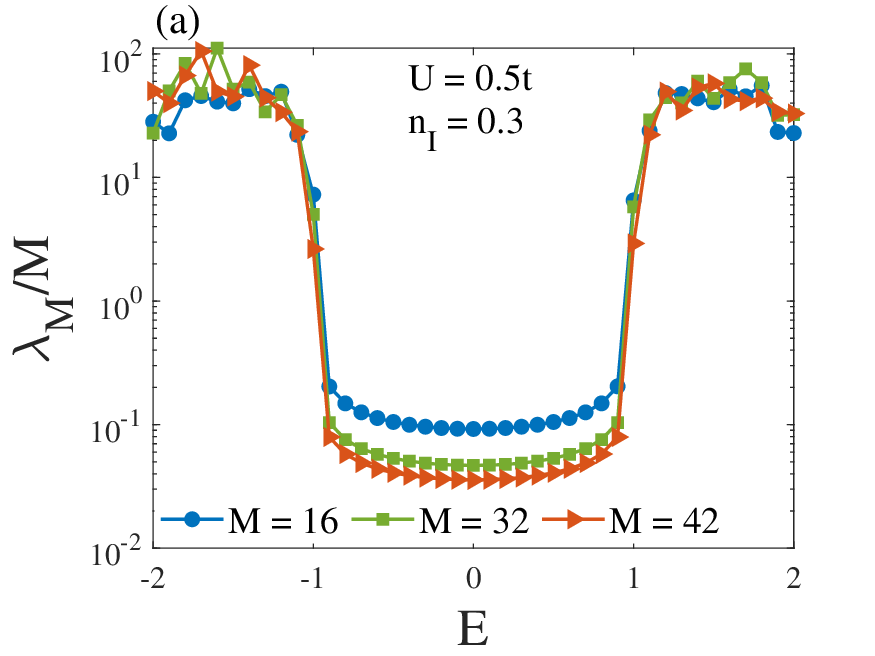}
\includegraphics[width=0.7\columnwidth]{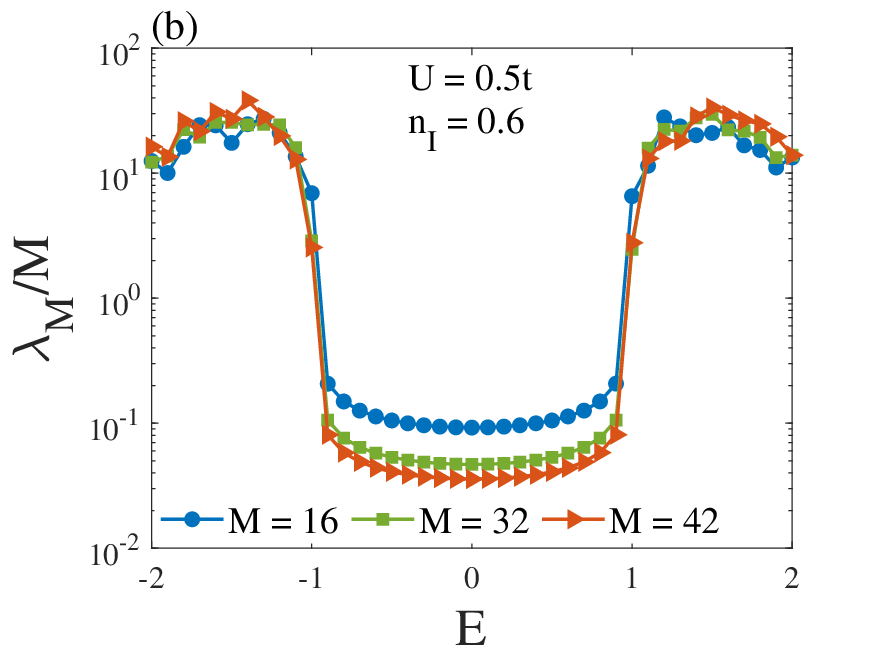}
\includegraphics[width=0.7\columnwidth]{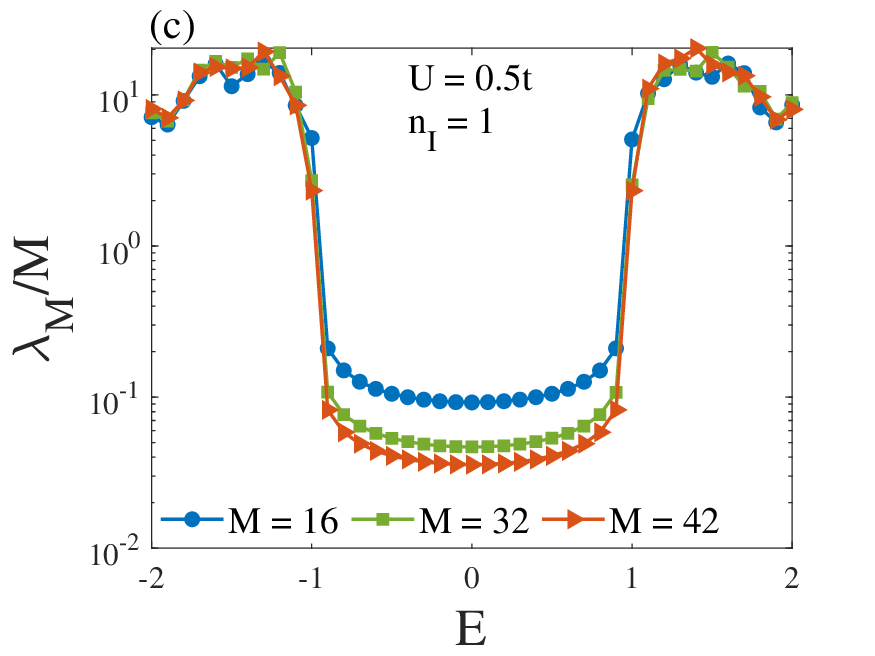}\\
\includegraphics[width=0.7\columnwidth]{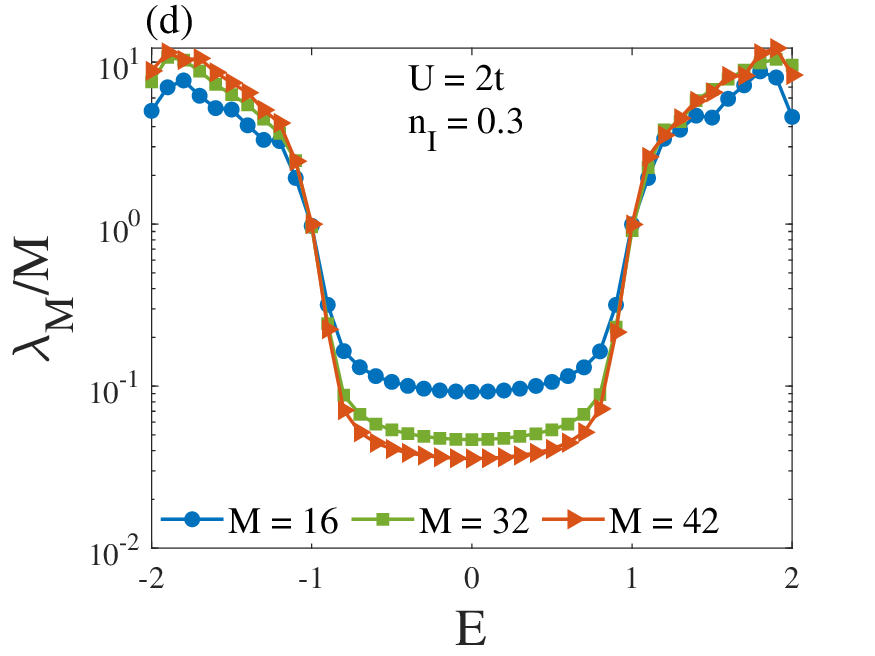}
\includegraphics[width=0.7\columnwidth]{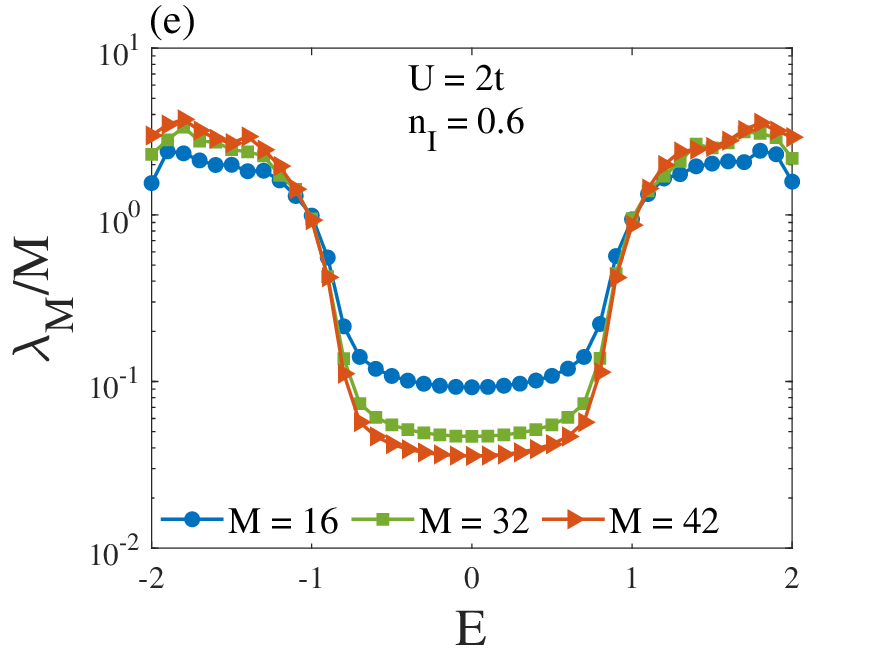}
\includegraphics[width=0.7\columnwidth]{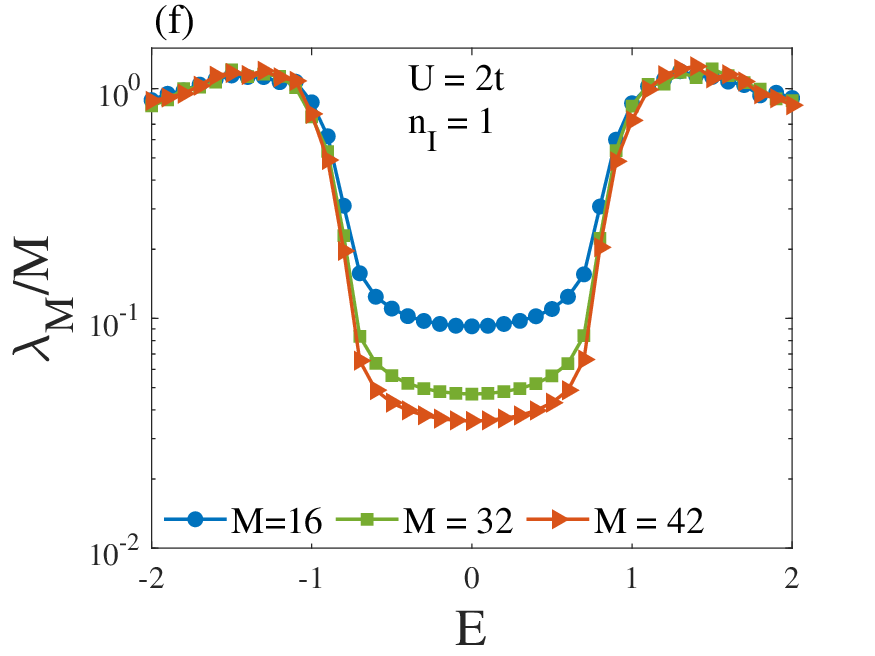}\\
\includegraphics[width=0.7\columnwidth]{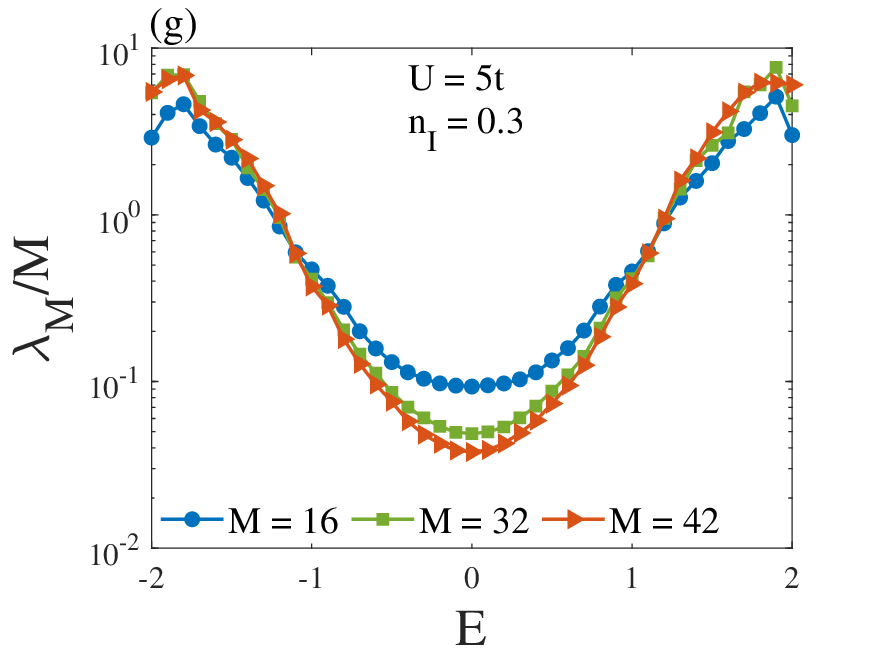}
\includegraphics[width=0.7\columnwidth]{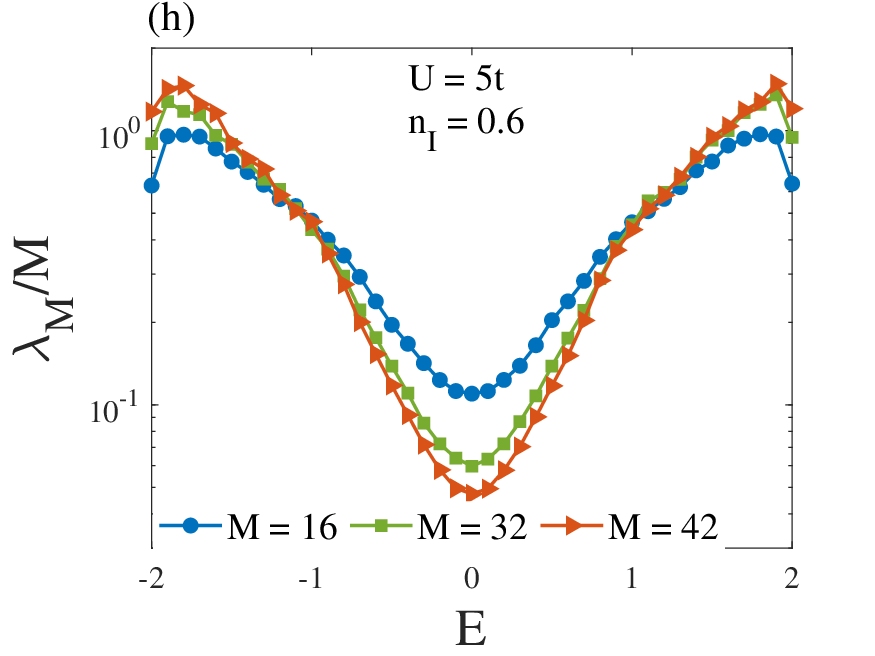}
\includegraphics[width=0.7\columnwidth]{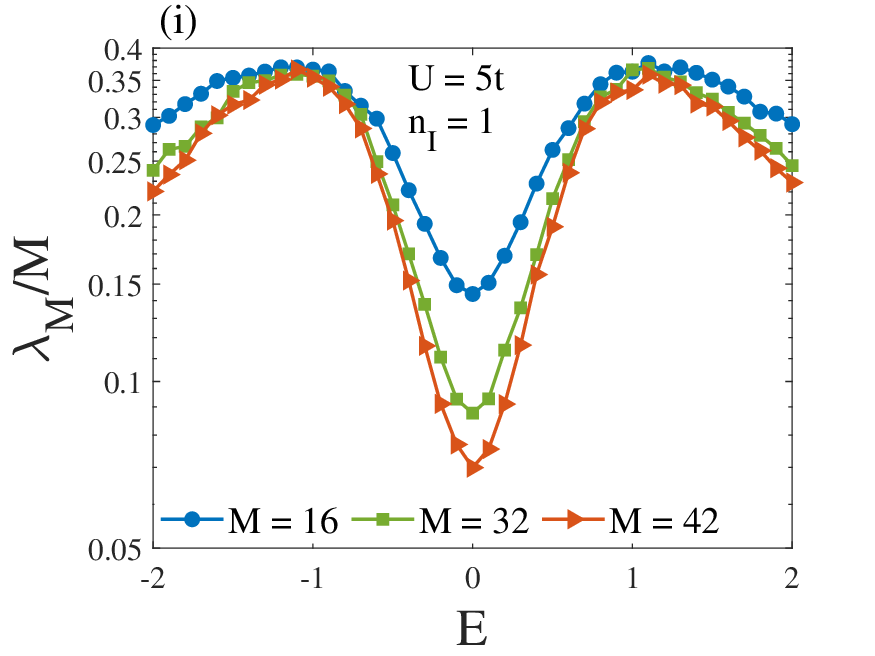}
\end{array}$
\caption{Normalized localization length $\Lambda_M=\frac{\lambda_M}M$ of the 2D HM  as a function of the energy $E$ expressed in unit of the NN integral $t$.
The results are shown for different ribbon sizes $M$. Calculations are done for $t_ 2 = 0.2t$, and $\Phi = \frac{\pi}2$. Each column (row) is for a given disorder concentration $n_I$ (disorder amplitude $U$).
The First, second and third columns correspond respectively to an impurity concentration $n_I=0.3$, $n_I=0.6$, and $n_I=1$ while the first, middle and last rows correspond to, respectively, $U=0.5 t$, $U=2 t$, and $U=5 t$.}
\label{loc-HM}
\end{figure*}
\begin{figure*}[hpbt] 
\centering
$\begin{array}{ccc}
\includegraphics[width=0.7\columnwidth]{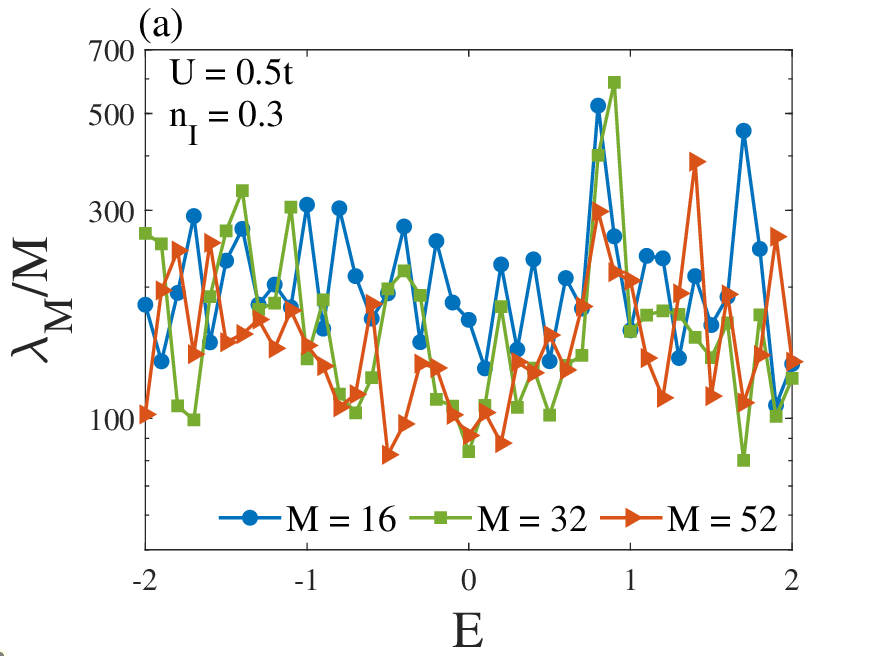}
\includegraphics[width=0.7\columnwidth]{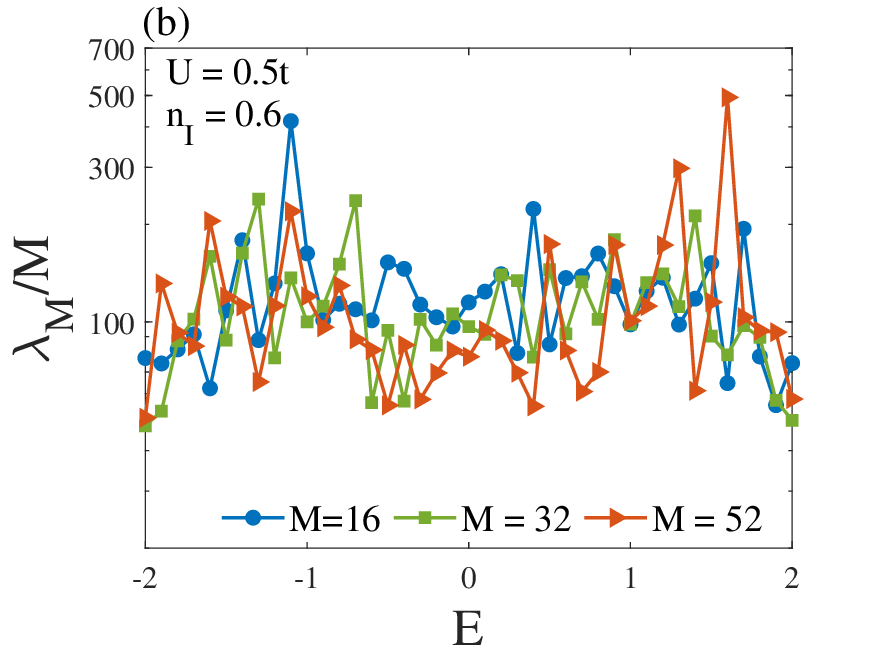}
\includegraphics[width=0.7\columnwidth]{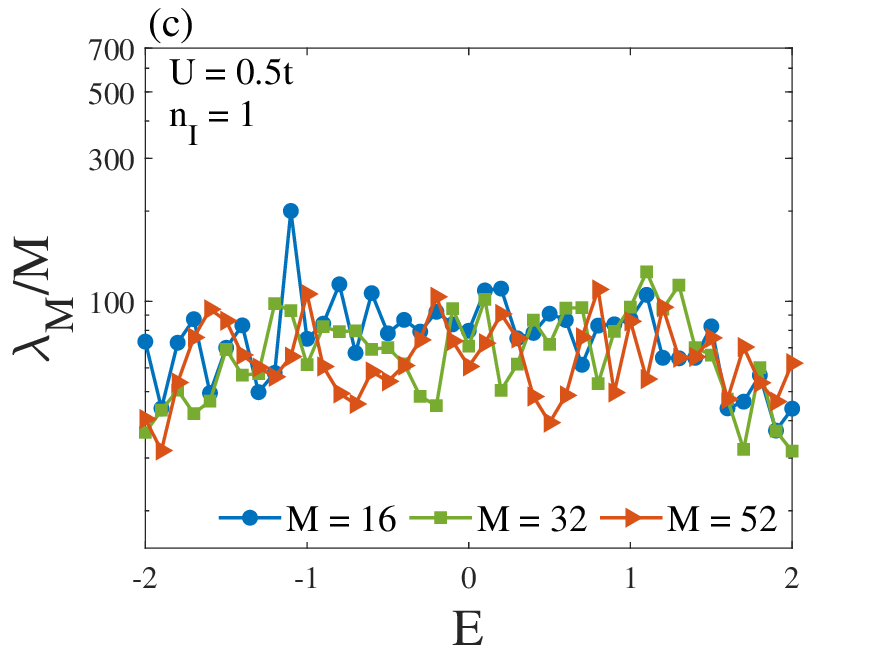}\\
\includegraphics[width=0.7\columnwidth]{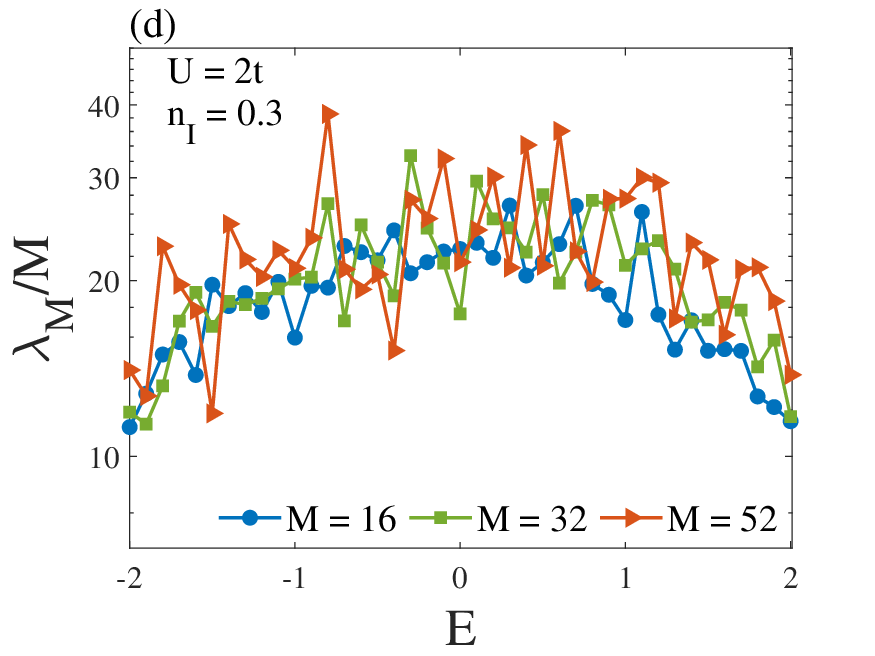}
\includegraphics[width=0.7\columnwidth]{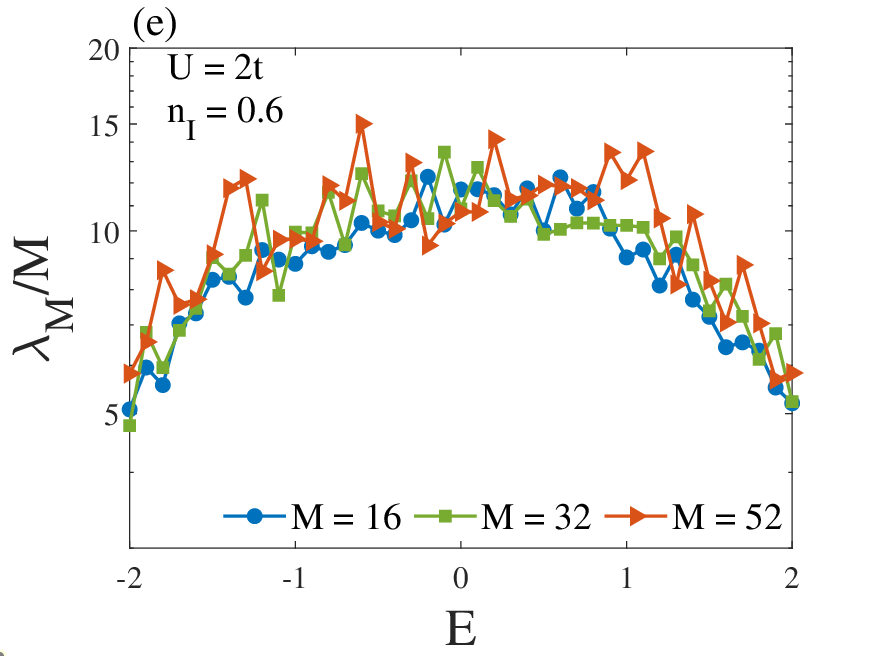}
\includegraphics[width=0.7\columnwidth]{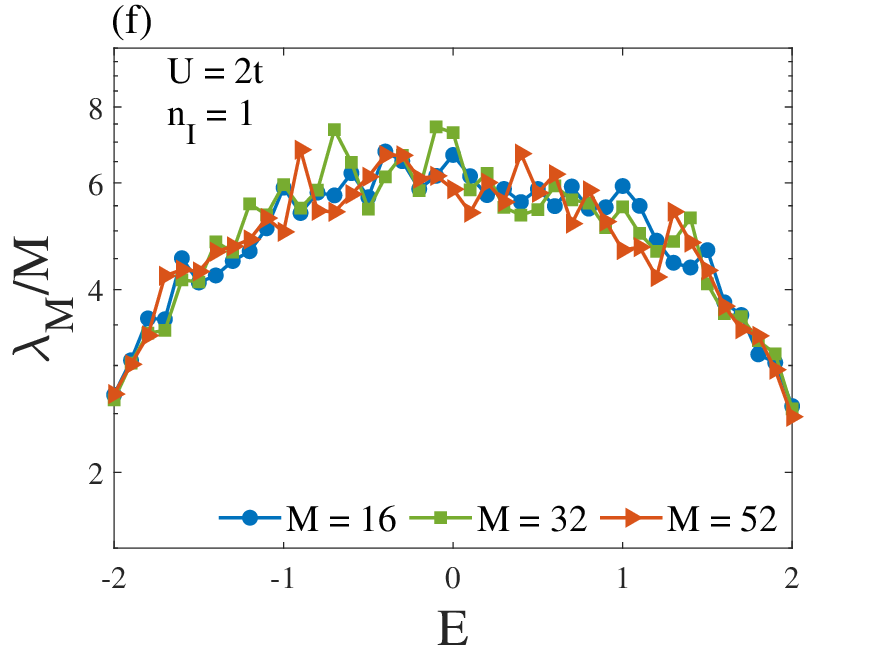}\\
\includegraphics[width=0.7\columnwidth]{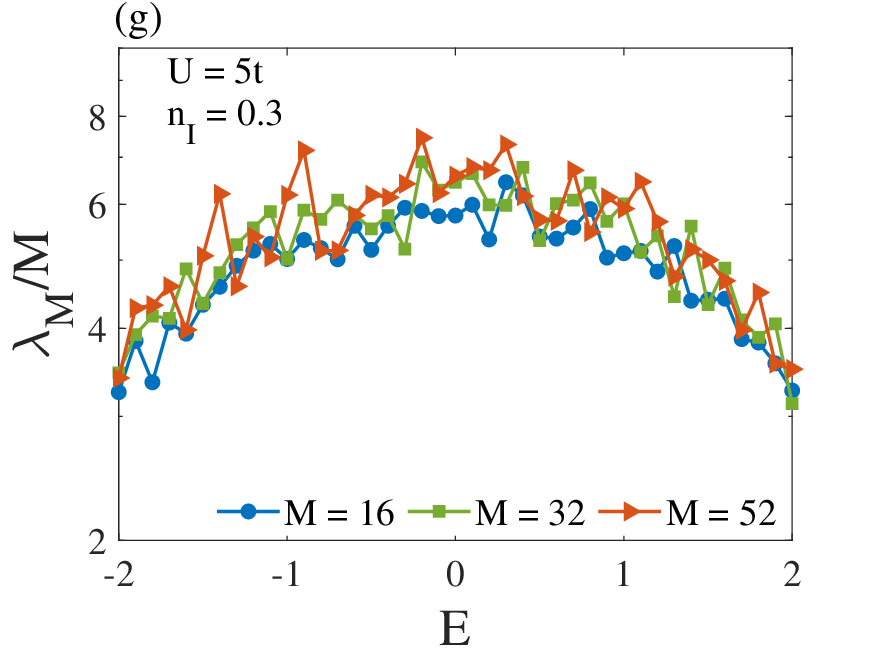}
\includegraphics[width=0.7\columnwidth]{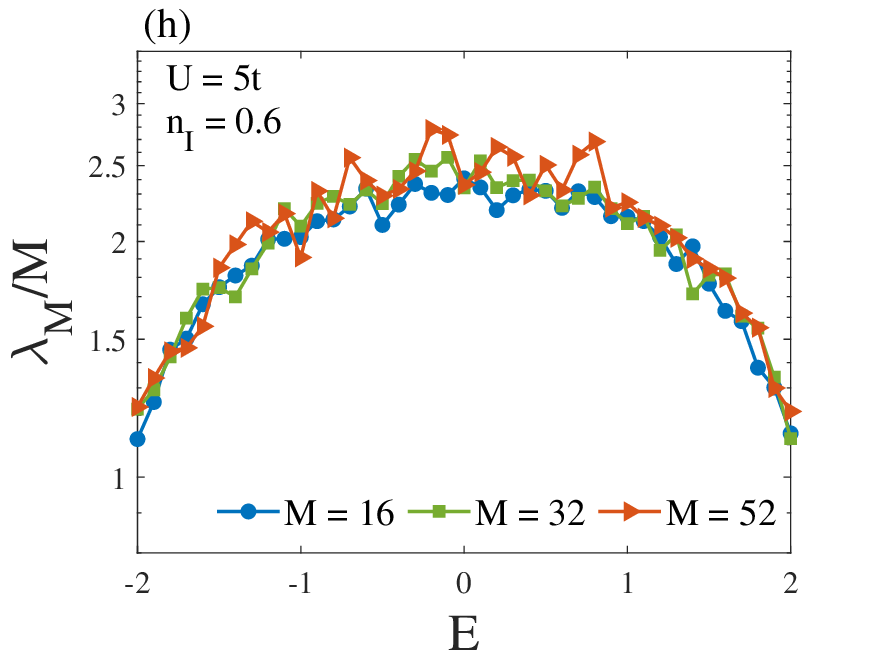}
\includegraphics[width=0.7\columnwidth]{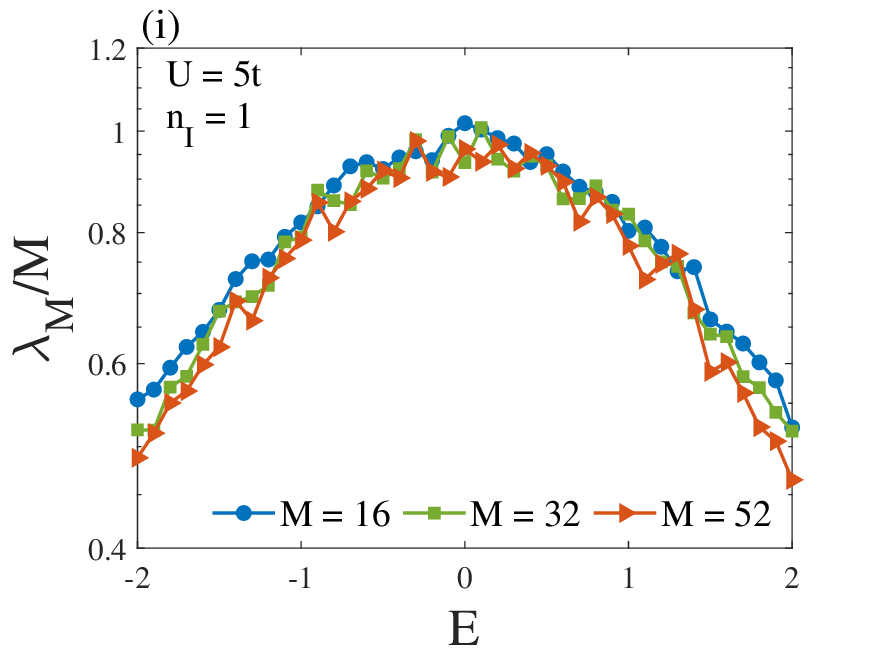}
\end{array}$
\caption{Normalized localization length $\Lambda_M=\frac{\lambda_M}M$ of the 2D mHM  as a function of the energy $E$ expressed in unit of the NN integral $t$.
The data are the same as in Fig.~\ref{loc-HM}.}
\label{loc-mHM}
\end{figure*}

We consider a graphene zigzag ribbon with $N_L$ unit cells along the zigzag periodic boundary and $M$ unit cells in the transverse direction, with $M\ll N_L$. The system can be descried by $N_L$ chains coupled by NN and NNN hopping processes, where each chain contains in total 2M atomic sites (Fig.~\ref{structure}).\

Based on the transfer matrix method (TMM)~\cite{TMM1,TMM2}, we computed the normalized localization length $\Lambda_M=\frac{\lambda_M}M$ of the HM and the mHM which are plot, respectively, in Figs.~\ref{loc-HM} and~\ref{loc-mHM}. $\lambda_M$ is the localization length defined as $\lambda_M=\gamma^{-1}$ and $\gamma$ is the Lyapunov exponent of the system.\

The behavior of $\Lambda_M$ with the system width $M$ gives insights into the localization properties of the wavefunction in the presence of disorder:
a decrease of $\Lambda_M$ with increasing $M$ is the signature of a disorder induced localization of the wavefunction. However, if $\Lambda_M$ increases or remains unchanged as  $M$ increases, then the wavefunction is extended.\
Fig.~\ref{loc-HM} shows that, as the ribbon width ($M$) increases, $\Lambda_M$ of the HM decreases, except at two energies where it remains unchanged. These energies correspond to the chiral edge states which keep their extended character along the sample terminations. These edge states survive even at strong disorder amplitude $U=5t$ and high impurity concentration $n_I=1$, where all the sites are subject to the random distributed impurities (Fig.~\ref{loc-HM} (i)). In this disorder configuration, $\Lambda_M$ decreases by increasing the system width $M$, for all the energies, except at two values corresponding to the chiral modes. This indicate that all the bulk states undergo an Anderson localization while the chiral states preserve their extended character.\

In the case of the mHM (Fig.~\ref{loc-mHM}), one cannot distinguish between the AC edge states and the bulk modes. Due to defect, the AC edge states can be coupled to the accompanying bulk states with which they propagate at the bulk boundaries~\cite{Franz,Mandal22}. 
In particular, at strong disorder amplitude (Fig.~\ref{loc-mHM} (c), (f) and (i)), $\Lambda_M$ decreases with increasing $M$ over all the energy spectrum, which means that the AC edge and the bulk states are localized by the disorder.\\

\section{Concluding remarks }
We addressed the robustness of the antichiral edge states (AC) occurring in the so-called modified Haldane model (mHM)~\cite{Franz}, describing a semi-metal with a broken TRS induced by a valley dependent scalar potential. We analyzed the behavior of the AC in the presence of Anderson disorder. We first discussed the reduced 1D mHM, which was reported~\cite{Franz} to have a topological character described by a momentum-dependent winding number. Our results show that this topological invariant is drastically suppressed even at small disorder amplitude. To avoid any misleading or incomplete conclusions drawn up from the reduced 1D model, we studied the localization behavior of the edge states of the HM and mHM in a ribbon geometry with a finite size. Using the transfer matrix method, we computed the localization lengths and showed that the AC edge states of the mHM are easily localized by disorder while the chiral ones remain robust even at relatively strong disorder amplitude. To sum up, we showed that the AC edge states are not robust against disorder regarding their coupling to the counter-propagating pseudo-bulk states. The non-vanishing conductance of the mHM ribbon, reported in Ref.~[\onlinecite{Franz}], is not necessarily a signature of the robustness of the AC edge states. It can be due to the pseudo-bulk states which are found to be peculiarly localized at the ribbon boundaries~\cite{Franz}
The fragile character of the AC edge states can be checked in photonic crystals ~\cite{Mandal}, electrical circuits~\cite{Yang21} and acoustic systems ~\cite{acoustic}. Our results can be used in the rapidly growing fields of topological photonics~\cite{Ozawa} and acoustics~\cite{topo-acous}, with antichiral propagating modes~\cite{JAP21,Chen20,Zhou,photonic22,surface-AC}, where lattice defects can be controlled to improve the robustness of the edge state transport.

\section{Acknowledgment }
We thank A. Cook and G. Lange for fruitful discussions and a critical reading of the manuscript. This work was supported by the Tunisian "Minist\`ere de l'Enseignement Sup\'erieur et de la Recherche Scientifique". M. M. is grateful for the financial support from the University of Porto. E. V. C. acknowledges partial support from FCT-Portugal through Grant No. UIDB/04650/2020. S. H. acknowledges the Max Planck Institute for the Physics of Complex Systems and the Institute for Theoretical Solid State Physics at IFW for kind hospitality and financial support.
\clearpage

%


\begin{thebibliography} {200}

%
\bibitem{Hasan-Rev} M. Z. Hasan and C. L. Kane, Rev. Mod. Phys. {\bf 82}, 3045 (2010).

\bibitem{Qi-Rev} X.-L. Qi and S.-C. Zhang, Rev. Mod. Phys. {\bf 83}, 1057 (2011).

\bibitem{Bansil} A. Bansil, H. Lin, and T. Das, Rev. Mod. Phys. {\bf 88}, 021004 (2016).

\bibitem{Haldane} F. D. M. Haldane, Phys. Rev. Lett. {\bf 61}, 2015 (1988).

\bibitem{AQHE-rev} K. He, Y. Wang and Q.-K. Xue, Annu. Rev. Condens. Matter Phys. {\bf 9}, 329 (2018).
%

\bibitem{Haldane2}  F. D. M. Haldane, and  S. Raghu, Phys. Rev. Lett. {\bf 100}, 013904 (2008).

\bibitem{Haldane3} S. Raghu, and F. D. M. Haldane, Phys. Rev. A {\bf 78}, 033834 (2008).

\bibitem{photonic} Z. Wang, Y. D. Chong, J. D. Joannopoulos, and M. Soljacic, Phys. Rev. Lett. {\bf 100}, 013905 (2008).

\bibitem{Niu} H. Jiang, Z. Qiao, H. Liu, and Q. Niu, Phys. Rev. B {\bf 85}, 045445 (2012).

\bibitem{Wang} Z. F. Wang, Z. Liu, and F. Liu, Phys. Rev. Lett. {\bf 110}, 196801 (2013).

\bibitem{Yang} Z. Yang, F. Gao, X. Shi, X. Lin, Z. Gao, Y. Chong, and B. Zhang,
Phys. Rev. Lett. {\bf 114}, 114301 (2015).

\bibitem{Khan} A. Khanikaev, R. Fleury, S. Mousavi, and A. Alu, Nat. Commun. {\bf 6}, 8260 (2015).

\bibitem{Ni} X. Ni, C. He, X.-C. Sun, X.-P. Liu, M.-H. Lu, L. Feng and Y.-F. Chen,   New J. Phys. {\bf 17} 053016 (2015).

\bibitem{kagome} G. Xu, B. Lian, and S.-C. Zhang, Phys. Rev. Lett. 115, 186802 (2015).


\bibitem{Kee} H.-S. Kim and H-Y Kee, npj Quantum Materials, {\bf 2}, 20 (2017).

\bibitem{Chern-gr} J. Zhang, B. Zhao, T. Zhou, Y. Xue, C. Ma, and Z. Yang, Phys. Rev. B {\bf 97}, 085401 (2018).
%

\bibitem{Chern-weyl} S. Howard, L. Jiao, Z. Wang, N. Morali, R. Batabyal, P. Kumar-Nag, N. Avraham, H. Beidenkopf, P. Vir, E. Liu, C. Shekhar, C. Felser, T.r Hughes and V. Madhavan, 
Nat. Commun. {\bf 12}, 4269 (2021).

\bibitem{Wang09} Z. Wang, Y. Chong, J. Joannopoulos, and M. Solja\u{c}i\'{c}, Nature {\bf 461}, 772 (2009).

\bibitem{cold} M. Mancini, G. Pagano, G. Cappellini, L. Livi, M. Rider, J. Catani, C. Sias, P. Zoller, M. Inguscio, M. Dalmonte, and L. Fallani, Science, {\bf 349} 1510 (2015).

\bibitem{acoustic} Y. Ding, Y. Peng, Y. Zhu, X. Fan, J. Yang, B. Liang, X. Zhu, X. Wan, and J. Cheng, Phys. Rev. Lett. {\bf 122}, 014302 (2019).

\bibitem{Rosen} I. T. Rosen, E. J. Fox, X. Kou, L. Pan, K. L. Wang and D. Goldhaber-Gordon, npj Quantum Mater. {\bf 2}, 69 (2017).

\bibitem{gr} X. Xi, J. Ma, S. Wan, C. -H. Dong, X. Sun, Sci. Adv. {\bf 7} eabe1398 (2021).


\bibitem{Franz} E. Colom\'es and M. Franz, Phys. Rev. Lett. {\bf 120}, 086603 (2018).

\bibitem{Mandal} S. Mandal, R. Ge, and T. C. H. Liew, Phys. Rev. B {\bf 99}, 115423 (2019).

\bibitem{Mandal22} R. Bao, S. Mandal, H. Xu, X. Xu, R. Banerjee and T. C. H. Liew, Phys. Rev. B {\bf 106}, 235310 (2022).

\bibitem{Chen20} J. Chen, W. Liang, and Z.-Y. Li, Phys. Rev. B {\bf 101}, 214102 (2020).

\bibitem{JAP21} L. Yu, H. Xue and B. Zhang, J. Appl. Phys. {\bf 129}, 235103 (2021). 

\bibitem{Denner} M. M. Denner, J. L. Lado, and O. Zilberberg, Phys. Rev. Research {\bf 2}, 043190 (2020).

\bibitem{Cheng} X. Cheng, J. Chen, L. Zhang, L. Xiao, and S. Jia, Phys. Rev. B {\bf 104}, L081401 (2021).

\bibitem{Bhowmick20} D. Bhowmick and P. Sengupta, Phys. Rev. B {\bf 101}, 195133 (2020).

\bibitem{Floquet} J. Wang, X. Ji, Z. Shi, Y. Zhang, H. Li, Y. Li, Y. Deng, K. Xie, arXiv:2302.05036.

\bibitem{Mannai} M. Manna\"i and S. Haddad, J. Phys.: Condens. Matter {\bf 32} 225501 (2020).

\bibitem{Roche} M. Vila, N. T. Hung, S. Roche, and R. Saito, Phys. Rev. B {\bf 99}, 161404(R) (2019).

\bibitem{supra} C. Wang, L. Zhang, P. Zhang, J. Song, and Y.-X. Li, Phys. Rev. B {\bf 101}, 045407 (2020).

\bibitem{Chen} X. -L. L\"{u}, J.-E. Ynag, and H. Chen, New J. Phys. {\bf 24}, 103021 (2022).

\bibitem{AC-mag} Z. Ji, J. Chen, and Z.-Y. Li, J. Appl. Phys. {\bf 133}, 140901 (2023).

\bibitem{helical} L. Xie, L. Jin, and Z. Song, Science Bulletin {\bf 68}, 255 (2023).

\bibitem{coexist} H. Wang, B. Xie, and W. Ren, Laser Photonics Rev. 2300764 (2023).

\bibitem{Marwa23} M. Manna\"{i}, J.-N. Fuchs, F. Pi\'echon, and S. Haddad, Phys. Rev. B {\bf 107}, 045117 (2023).

\bibitem{heteroHM} J.Chen and Z.-Y. Li, Phys. Rev. Lett. {\bf 128}, 257401  (2022).

\bibitem{Zhou} P. Zhou, G.-G. Liu, Y. Yang, Y.-H. Hu, S. Ma, H. Xue, Q. Wang, L. Deng, and B. Zhang, Phys. Rev. Lett. {\bf 125}, 263603 (2020).

\bibitem{photonic22} J. Chen and Z.-Y. Li, Opto-Electron Sci. {\bf 1}, 220001 (2022).

\bibitem{Yang21} Y. Yang, D. Zhu, Z. Hang, and Y. Chong, Sci. China Phys. Mech. Astron. {\bf 64}, 257011 (2021).

\bibitem{Weyl} X. Xi, B. Yan, L. Yang, Y. Meng, Z.-X. Zhu, J.-M. Chen, Z. Wang, P. Zhou, P. P. Shum, Y. Yang, H. Chen, S. Mandal, G.-G. Liu, B. Zhang and Z.Gao, Nat. Commun. {\bf 14}, 1991 (2023). 

\bibitem{surface-AC} J. W. Liu , F. L. Shi, K. Shen, X. D. Chen, K. Chen, W. J. Chen, and J. W. Dong, Nat Commun. {\bf 14} 2027 (2023).


\bibitem{Gr-cond} I. Kleftogiannis, I. Amanatidis, and V. A. Gopar, Phys. Rev.B {\bf 88}, 205414 (2013).

\bibitem{AZ} A. Altland, and M. R. Zirnbauer, Phys. Rev. B {\bf 55}, 1142 (1997); C.-K. Chiu, J. C. Y. Teo, A. P. Schnyder, and S. Ryu, Rev. Mod. Phys. {\bf 88}, 035005 (2016).

\bibitem{SSH-HM} L. Li, Z. Xu, and S. Chen, Phys. Rev. B {\bf 89}, 085111 (2014).

\bibitem{SSH} W. P. Su, J. R. Schreiffer, and A. J. Heeger, Phys. Rev. Lett. {\bf 42}, 1698 (1979).

\bibitem{Phil} P. W. Anderson, Phys. Rev. {\bf 109}, 1492 (1958).

\bibitem{Bena} C. Bena and G. Montambaux, New J. Phys. {\bf 11}, 095003 (2009).

\bibitem{JN} J.-N. Fuchs and F. Piéchon, Phys. Rev. B {\bf 104}, 235428 (2021).

\bibitem{Vanderbilt} D. Vanderbilt, {\it Berry phase in electronic structure theory: electric polarization, orbital magnetization and topological insulators} (Cambridge university press, 2018).

\bibitem{Cayssol} J. Cayssol and J.-N. Fuchs, J. Phys. Mater. {\bf 4} 034007 (2021).

\bibitem{CM} Y. -F. Zhang, Y.-Y. Yang, Y. Ju, L. Sheng, R. Shen, D.-N. Sheng and D.-Y. Xing, Chinese Phys. B {\bf 22} 117312 (2013).


\bibitem{SSH-IPR} F. Munoz, F. Pinilla, J. Mella and M. I. Molina, Scientific Reports {\bf 8}, 17330 (2018).

\bibitem{IPR} A. Mart\'{i}n Pend\'{a}s, F. Mu\~{n}oz, C. Cardenas,  and J. Contreras-Garc\'{\i}a, Molecules {\bf 26}, 2965 (2021).

\bibitem{TMM1} A. MacKinnon, B. Kramer, Z. Physik B - Condensed Matter {\bf 53}, 1 (1983).

\bibitem{TMM2} B. Kramer and A. MacKinnon,  Rep. Prog. Phys. {\bf 56}, 1469 (1993).

\bibitem{Ozawa} T. Ozawa, H. M. Price, A. Amo, N. Goldman, M. Hafezi, L. Lu, M. C. Rechtsman, D. Schuster, J. Simon, O. Zilberberg, and I. Carusotto, Rev. Mod. Phys {\bf 91}, 015006 (2019).

\bibitem{topo-acous} H. Xue, Y. Yang, and B. Zhang, Nat. Rev. Mater. {\bf 7}, 974 (2022). 
\end{thebibliography}
\end{document}